\documentclass[10pt]{iopart}
\expandafter\let\csname equation*\endcsname=\relax
\expandafter\let\csname endequation*\endcsname=\relax
\usepackage{graphicx}
\usepackage{bm}
\usepackage{amsmath}
\usepackage{amssymb}
\usepackage{physics}
\usepackage[colorlinks=true,linkcolor=blue,urlcolor=blue,citecolor=blue]{hyperref}
\usepackage[usenames,svgnames]{xcolor}
\usepackage{makecell}
\usepackage[normalem]{ulem}
\usepackage{cite}
\usepackage[left=1in,right=1in]{geometry}

\newcommand{\Hop}{\hat{H}}
\newcommand{\Himp}{\hat{H}_{\rm imp}}
\newcommand{\Uimp}{\hat{U}_{\rm imp}}
\newcommand{\gHimp}{\mathcal{H}_{\rm imp}}

\newcommand{\Hbath}{\hat{H}_{\rm bath}}
\newcommand{\Hhyb}{\hat{H}_{\rm hyb}}

\newcommand{\aop}{\hat{a}}
\newcommand{\adop}{\hat{a}^{\dagger}}

\newcommand{\cop}{\hat{c}}
\newcommand{\cdop}{\hat{c}^{\dagger}}

\newcommand{\Zimp}{Z_{{\rm imp}}}
\newcommand{\mea}{\mathcal{D}}
\newcommand{\gK}{\mathcal{K}}
\newcommand{\gI}{\mathcal{I}}
\newcommand{\bolda}{\bm{a}}
\newcommand{\boldabar}{\bar{\bm{a}}}
\newcommand{\abar}{\bar{a}}

\newcommand{\gA}{\mathcal{A}}
\newcommand{\gB}{\mathcal{B}}

\newcommand{\propagator}{\mathcal{G}}

\newcommand{\timecost}{\mathcal{C}^t}
\newcommand{\spacecost}{\mathcal{C}^s}
\newcommand{\hnu}{Key Laboratory of Low-Dimensional Quantum Structures and Quantum Control of Ministry of Education, Department of Physics and Synergetic Innovation Center for Quantum Effects and Applications, Hunan Normal University, Changsha 410081, China
}

\begin{document}

\title{Grassmann Time-Evolving Matrix Product Operators for Equilibrium Quantum Impurity Problems}
\date{\today}

\author{Ruofan Chen$^1$, Xiansong Xu$^{2,1}$, Chu Guo$^{3,*}$}
%\email{guochu604b@gmail.com}

\address{$^1$ College of Physics and Electronic Engineering, and Center for Computational Sciences, Sichuan Normal University, Chengdu 610068, China}
\address{$^2$ Science, Mathematics and Technology Cluster, Singapore University of Technology and Design, 8 Somapah Road, Singapore 487372}
\address{$^3$ \hnu}
\address{$^*$ Email: \mailto{guochu604b@gmail.com}}


\begin{abstract}
Tensor-network-based methods are promising candidates to solve quantum impurity problems. They are free of sampling noises and the sign problem compared to state-of-the-art continuous-time quantum Monte Carlo methods. Recent progress made in tensor-network-based impurity solvers is to use the Feynman-Vernon influence functional to integrate out the bath analytically, retaining only the impurity dynamics and representing it compactly as a matrix product state. 
The recently proposed Grassmann time-evolving matrix product operator (GTEMPO) method is one of the representative methods in this direction.
In this work, we systematically study the performance of GTEMPO in solving equilibrium quantum impurity problems at a finite temperature with a semicircular spectrum density of the bath. Our results show that its computational cost would generally increase as the temperature goes down and scale exponentially with the number of orbitals. In particular, the single-orbital Anderson impurity model can be efficiently solved with this method, for two orbitals we estimate that one could possibly reach inverse temperature $\beta\approx 20$ if high-performance computing techniques are utilized, while beyond that only very high-temperature regimes can be reached in the current formalism. Our work paves the way to apply GTEMPO as an imaginary-time impurity solver.
\end{abstract}

\maketitle

%%%%% introduction

\section{Introduction}

Quantum impurity problems (QIPs) are prototypical models for studying open quantum systems and strongly correlated phenomena~\cite{Wilson1975,hewson1993-the}. They are also the central ingredients for quantum embedding theories such as the dynamical mean field theory~\cite{GeorgesRozenberg1996}. However, the strongly correlated nature of the problem and the exponential growth of the degrees of freedom pose great challenges for their accurate numerical solutions. 

In this work, we focus on the equilibrium scenario where the impurity and the bath are in thermal equilibrium with a finite temperature and aim to calculate the Matsubara Green's function.
For such problems, the current method of choice is the class of continuous-time quantum Monte Carlo (CTQMC) methods, due to their efficiency and accuracy~\cite{RubtsovLichtenstein2004,RubtsovLichtenstein2005,WernerMillis2006,GullTroyer2008,ChanMillis2009,haule2010-dynamical,GullWerner2011,huang2014-electronic,lu2016-pressure,yue2021-pairing}. These methods are free of bath discretization errors as the baths are integrated out exactly. Moreover, they are also free of time discretization errors and can cope with up to five orbitals. The possible drawbacks of these methods are that sampling noises are unavoidable as a general feature of Monte Carlo methods, and they could suffer from the sign problem~\cite{TroyerWiese2005}. In addition, since they mostly work in the imaginary-time axis, one needs to perform analytical continuation to obtain the spectral function, which is a numerically ill-posed operation~\cite{FeiGull2021}.

Tensor-network-based methods are a class of methods that can in principle solve these drawbacks of CTQMC methods and also be scalable to multi-orbital QIPs. Various tensor-network-based impurity solvers have been proposed to date.
For example, methods by representing the impurity-bath wave function as a matrix product state (MPS) have been applied to study equilibrium QIPs in both the real-time~\cite{WolfSchollwock2014b,GanahlEvertz2014,GanahlVerstraete2015} and imaginary-time axis~\cite{WolfSchollwock2015}. These methods usually focus on the zero temperature scenario due to the difficulty of representing a finite-temperature bath state as an MPS. We note the recent works that overcome this difficulty using the thermofield transformations~\cite{TakahashiUmezawa1996,InesBanuls2015} to avoid explicit preparation of thermal bath states~\cite{KohnSantoro2021,KohnSantoro2022}, but are limited to a single orbital. The numerical renormalization group method~\cite{BullaPruschke2008} is able to deal with finite-temperature baths and has been demonstrated up to three orbitals~\cite{MitchellBulla2014,StadlerWeichselbaum2015,HorvatMravlje2016,KuglerGeorges2020}, which, however, mostly focuses on the low-energy spectrum and could be less accurate in the high-energy part. Generally, these tensor-network-based methods still cannot match CTQMC methods in terms of efficiency, accuracy or flexibility. This is mainly because they need to explicitly deal with the bath, leading to bath discretization errors as well as higher numerical costs.

In the last five years, important progress has been made in tensor-network-based impurity solvers by using the Feynman-Vernon influence functional (IF)~\cite{FeynmanVernon1963} to integrate out the bath analytically, which avoids explicitly treatment on the bath. 
This idea has been first implemented in bosonic systems under the name of the time-evolving matrix product operators (TEMPO)~\cite{StrathearnLovett2018}, which has achieved unprecedented accuracy and efficiency in studying the non-equilibrium dynamics for bosonic impurity problems~\cite{joergensen2019-exploiting,popovic2021-quantum,fux2021-efficient,gribben2021-using,otterpohl2022-hidden,gribben2022-exact}, and then in fermionic systems~\cite{ThoennissAbanin2023a,ThoennissAbanin2023b,NgReichman2023,KlossAbanin2023} (the latter methods will be referred to as the tensor network IF methods in the following). In our previous work, we propose a Grassmann time-evolving matrix product operator method which is a full fermionic analog of the TEMPO method~\cite{ChenGuo2023}. Compared to the tensor network IF methods, GTEMPO has several advantages originating from TEMPO: 1) it is universal for both real- and imaginary-time (or even mixed-time) calculations and is transparent to implement as it only depends on the Grassmann path integral (PI). In comparison, the tensor network IF method requires further additional steps to convert the Grassmann PI back into fermionic expectation value calculations; 2) it efficiently constructs the IF as a Grassmann MPS (GMPS), referred to as the MPS-IF, using a series of GMPS multiplications. Meanwhile, the tensor network IF method uses additional transformations to build the IF as an MPS which also introduces new hyperparameters into the algorithm. For non-equilibrium dynamics of the single-orbital Anderson impurity model (AIM), it has been demonstrated that GTEMPO can reach higher precision with lower numerical cost compared to the tensor network IF methods~\cite{ChenGuo2023}.

Although the GTEMPO method is universal for very broad types of QIPs, the current study is limited to the non-equilibrium scenario~\cite{ChenGuo2023}.
In this work, we systematically study the efficiency of the GTEMPO method for equilibrium QIPs at finite temperatures, focusing on a semicircular spectrum of the bath which is commonly used in literature.
% The crucial factor that affects the performance of it is the bond dimension of the MPS-IF. 
The article is organized as follows. In Sec.~\ref{sec:method}, we describe the method in detail, as well as techniques specific to the finite-temperature scenario to accelerate its performance. In Sec.~\ref{sec:toulouse}, we show the accuracy of the method for the analytically solvable case of non-interacting fermions and numerically study the growth of the bond dimension of the MPS-IF. It allows us to accurately estimate the computational costs of multi-orbital AIMs. In Sec.~\ref{sec:oneorbitals}, we study the single-orbital AIM for different interaction strengths and compare our results to CTQMC calculations. In Sec.~\ref{sec:moreorbitals}, we present high-temperature calculations for higher-orbital AIMs which show the flexibility as well as the current limitation of the method. 

% Our results show that the single-orbital Anderson model is well within reach for GTEMPO. For two-orbital QIPs it could reach inverse temperature $\beta\approx 40$ if high performance computing techniques are utilized. Beyond that only very high temperature regimes may be reached in its current formalism.

\section{GTEMPO for equilibrium QIPs}\label{sec:method}
In this section, we describe the major steps of the GTEMPO method when working in the imaginary-time axis to solve equilibrium QIPs at finite temperatures. In addition, we present techniques specific to such a scenario to accelerate the numerical performance.

\subsection{Description of the Hamiltonian}

The Hamiltonian for a general QIP can be written as 
\begin{align}
\Hop = \Himp + \Hbath + \Hhyb,
\end{align}
where $\Himp$ denotes the impurity Hamiltonian, $\Hbath$ denotes the bath Hamiltonian, and $\Hhyb$ denotes the hybridization Hamiltonian between the impurity and the bath.
% In this work we consider 
The impurity Hamiltonian $\Himp$ can be generally written in the form
\begin{align}\label{eq:Himp}
\Himp = \sum_{p, q} t_{p, q} \adop_{p}\aop_{q} + \sum_{p,q,r,s} v_{p,q,r,s} \adop_p\adop_q\aop_r\aop_s ,
\end{align}
where $p,q,r,s$ are fermion flavors that contain both the spin and orbital indices. The first term on the right-hand side of Eq.(\ref{eq:Himp}) is the tunneling term and the second is the interaction term.
$\Hbath$ describes flavors of non-interacting fermions, which can be written as
\begin{align}
\Hbath = \sum_{p, k} \varepsilon_{p, k} \cdop_{p, k} \cop_{p, k}, 
\end{align}
where $k$ labels the momentum, and $\varepsilon_{p, k}$ is the corresponding energy. The number of flavors in $\Hbath$ is exactly the same as that in $\Himp$. In this work, we consider $\Hhyb$ in the form of linear coupling between the impurity and the bath
\begin{align}
\Hhyb = \sum_{p, k} V_{p, k}\left(\adop_{p, k} \cop_{p, k} + \cdop_{p, k}\aop_{p, k} \right), 
\end{align}
where $V_{p, k}$ is the hybridization strength. Throughout this work we assume $\Hbath$ and $\Hhyb$ to be independent of the flavors, namely $\varepsilon_{p,k} = \varepsilon_k$ and $V_{p, k} = V_k$, for simplicity of notations, but we note that GTEMPO can be straightforwardly applied to more general cases with no additional efforts (as long as $\Hhyb$ takes the form of linear coupling).

\subsection{Grassmann path integral on the imaginary-time axis}\label{sec:PI}
For equilibrium QIPs at inverse temperature $\beta$, the impurity partition function $\Zimp(\beta) = \Tr e^{-\beta\Hop}/\Tr e^{-\beta\Hbath}$ (we assume that the bath has zero chemical potential throughout this work) can be written as a PI over imaginary-time Grassmann trajectories~\cite{kamenev2009-keldysh,negele1998-quantum}:
\begin{align}\label{eq:PI}
  \Zimp(\beta)=\int\mea[\boldabar,\bolda]\gK[\boldabar,\bolda]\sum_p \gI[\boldabar_p,\bolda_p],
\end{align}
where $\boldabar_p=\{\abar_p(\tau)\}$ and $\bolda_p=\{a_p(\tau)\}$ are Grassmann trajectories for flavor $p$ over the continuous imaginary-time interval $[0, \beta]$, $\boldabar = \{\boldabar_p, \boldabar_q, \cdots\}$ and $\bolda = \{\bolda_p, \bolda_q, \cdots\}$ are Grassmann trajectories for all flavors. 
The measure is
\begin{align}
  \mea[\boldabar,\bolda]=\prod_{p, \tau}\dd\abar_p(\tau)\dd a_p(\tau)e^{-\abar_p(\tau)a_p(\tau)}.
\end{align}
$\gK[\boldabar,\bolda]$ encodes the bare impurity dynamics which only depends on $\Himp$.
$\gI[\boldabar_p,\bolda_p]$ is the IF for flavor $p$, which can be written as
\begin{align}
  \label{eq:IF}
  \gI[\boldabar_p,\bolda_p]=e^{-\int_0^{\beta}\dd\tau\int_0^{\beta}\dd\tau'\abar_p(\tau)\Delta(\tau,\tau')a_p(\tau')}.
\end{align}
The bath effect is fully characterized by the hybridization function $\Delta(\tau,\tau')$, which can be computed as
\begin{align}\label{eq:hybridization-function}
  \Delta(\tau,\tau')=\int \dd\varepsilon J(\varepsilon)D_{\varepsilon}(\tau,\tau').
\end{align}
Here $J(\varepsilon)=\sum_kV^2_k\delta(\varepsilon-\varepsilon_k)$ is
the bath spectrum density,
$D_{\varepsilon}(\tau,\tau')$
is the free bath Matsubara Green's function:
\begin{align}\label{eq:free-bath-D}
  D_{\varepsilon}(\tau,\tau')=\expval*{T_{\tau}\cop_{\varepsilon}(\tau)\cdop_{\varepsilon}(\tau')}_0 =[\Theta(\tau-\tau')-n(\varepsilon)]e^{-\varepsilon(\tau-\tau')},
\end{align}
where $n(\varepsilon)=(e^{\beta\varepsilon}+1)^{-1}$ is the
Fermi-Dirac distribution and $\Theta$ is the Heaviside step
function. Throughout this work we will consider a semicircular spectrum (this type of spectrum was also considered in, e.g., Refs. \cite{lauchli2009-krylov,wolf2014-solving,KohnSantoro2021,NgReichman2023})
\begin{align}
J(\varepsilon)=\frac{1}{\pi}D\sqrt{1-(\varepsilon/D)^2},
\end{align}
and set the half bandwidth $D=1$ as the unit.

To numerically evaluate the PI in Eq.\eqref{eq:PI}, we split $\beta=(M-1)\delta\tau$
and then the continuous trajectories $\{\abar_p(\tau)\},\{a_p(\tau)\}$ are
discretized as $\boldabar_p=\{\abar_{p,M},\ldots,\abar_{p,1}\}$ and
$\bolda_p=\{a_{p,M},\ldots,a_{p,1}\}$ accordingly (we still use the same notation as the continuous case to reduce the number of notations). 
In addition, we denote $\boldabar_{, j}=\{\boldabar_{p, j}, \boldabar_{q, j}, \dots\}$ and $\bolda_{, j}=\{\bolda_{p, j}, \bolda_{q, j}, \dots\}$ as the Grassmann variables (GVs) for all the flavors at discrete time step $j$.
Under these notations, the double
integral in the IF in Eq.\eqref{eq:IF} can be
discretized via the QuaPI method
\cite{makarov1994-path,makri1995-numerical,dattani2012-analytic} as
\begin{align}
\int_0^{\beta}\dd\tau\int_0^{\beta}\dd\tau' \abar_p(\tau)\Delta(\tau,\tau')a_p(\tau') \approx\sum_{i,j=1}^{M-1}\abar_{p,i}\Delta_{ij}a_{p,j},
\end{align}
where $\Delta_{ij}$ is the hybridization matrix which can be calculated as 
\begin{align}
  \Delta_{ij}=\int_{i\delta\tau}^{(i+1)\delta\tau}\dd\tau\int_{j\delta\tau}^{(j+1)\delta\tau}\dd\tau'
  \Delta(\tau,\tau').
\end{align}
We note that as compared to the non-equilibrium case where there are two branches in the real-time axis (forward and backward branches)~\cite{ChenGuo2023}, there is only one branch in the imaginary-time axis in the equilibrium case. As a result, the total number of GVs is reduced by half for the same $M$, and there is only one hybridization matrix $\Delta_{ij}$ after discretization.

The discretized bare impurity dynamics $\gK[\boldabar,\bolda]$ can be evaluated as 
\begin{align}
  \gK[\boldabar,\bolda]=\mel{-\bolda_{,1}}{\Uimp}{\bolda_{,M}}\cdots\mel{\bolda_{,2}}{\Uimp}{\bolda_{,1}},
\end{align}
where we have used $\Uimp = e^{-\delta\tau\Himp}$. Here the boundary condition, which corresponds to the final trace when evaluating $\Zimp(\beta)$ in Eq.(\ref{eq:PI}), is handled by GVs $\boldabar_{,1}$ and $\bolda_{,1}$. However in this
situation, the GVs $\boldabar_{,1},\bolda_{,1}$ do not belong to the same time
step: $\boldabar_{,1}$ live at $\tau=\beta$ while $\bolda_{,1}$ live at $\tau=0$, which 
is inconvenient for QuaPI. Thus we introduce extra GVs $\boldabar_{,0}$ and $\bolda_{,0}$ to solely take care of the boundary condition, and equivalently write $\gK[\boldabar,\bolda]$ as
% \begin{align}\label{eq:K-definition}
% \gK[\boldabar,\bolda]=&\braket{-\bolda_{,0}}{\bolda_{,M}}\mel{\bolda_{,M}}{\Uimp(\delta\tau)}{\bolda_{,M-1}} \times \nonumber \\
%     &\cdots\mel{\bolda_{,2}}{\Uimp(\delta\tau)}{\bolda_{,1}}\braket{\bolda_{,1}}{\bolda_{,0}}.
% \end{align}
\begin{align}\label{eq:K}
\gK[\boldabar,\bolda]=&\braket{-\bolda_{,0}}{\bolda_{,M}}\propagator_{M, M-1} \cdots \propagator_{2,1} \braket{\bolda_{,1}}{\bolda_{,0}}.
\end{align}
Now the first and last terms on the right-hand side of Eq.(\ref{eq:K}) are responsible for the boundary condition, and each term in the middle is a propagator from time step $j-1$ to $j$, defined as $\propagator_{j, j-1} := \mel{\bolda_{,j}}{\Uimp}{\bolda_{,j-1}}$.
Generally, a first-order expression of $\propagator_{j, j-1}$ is
\begin{align}\label{eq:G}
\propagator_{j, j-1} \approx \propagator_{j, j-1}^{(1)} := e^{\boldabar_{,j}\bolda_{,j-1}-\delta\tau\gHimp(\boldabar_{,j}, \bolda_{,j-1})} ,
\end{align}
where the term $\gHimp(\boldabar_{,j}, \bolda_{,j-1})$ in the exponent is simply obtained by replacing in Eq.(\ref{eq:Himp}) with $\adop_p \rightarrow \abar_{p,j}$ and $\aop_p\rightarrow a_{p,j-1}$. However, a first-order expression of the propagator would often result in very large numerical errors (See Fig.~\ref{fig:multiorbital1} for example). For simple models such as the single-orbital AIM, one could easily obtain the exact analytical expression of the propagator. In general cases, calculating the exact analytical expression for $\propagator_{j, j-1}$ is tedious and not scalable. Nevertheless, we propose the following trick which can calculate numerically accurate propagators to arbitrary precision with very little numerical cost. The idea is to break $\delta\tau$ into $m$ smaller time steps with $\delta\tau' = \delta\tau/m$. Correspondingly, one inserts $m-1$ pairs of GVs between $\bolda_{,j-1}$ and $\bolda_{,j}$, denoted as $\bolda_{, j, l}$ and $\boldabar_{, j, l}$ with $1\leq l\leq m-1$ . 
Then one can calculate a more precise $\propagator_{j, j-1}$ by starting from a refined discretization with step size $\delta\tau'$ and then integrating out the intermediate GVs, that is,
\begin{align}\label{eq:Gacc}
\propagator_{j, j-1} = \int \prod_{p, 1\leq l\leq m-1} \dd \abar_{p,j, l}\dd a_{p, j, l} e^{-\abar_{p, j, l}a_{p, j, l}}  
\mel{\bolda_{,j}}{\Uimp'}{\bolda_{,j, m-1}} \cdots 
\mel{\bolda_{,j, 2}}{\Uimp'}{\bolda_{,j,1}} \mel{\bolda_{,j, 1}}{\Uimp'}{\bolda_{,j-1}}, 
\end{align}
where the propagator $\Uimp'=e^{-\delta\tau'\Himp}$ in the integrand can be approximated by their first-order expressions as in Eq.(\ref{eq:G}).
With this trick, the numerical error in calculating $\propagator_{j, j-1}$ decreases from $O(\delta\tau)$ in Eq.(\ref{eq:G}) to $O(\delta\tau')$ in Eq.(\ref{eq:Gacc}). For time-independent $\Himp$, 
the propagator is time-translationally invariant. Therefore Eq.(\ref{eq:Gacc}) only needs to be evaluated once and then can be used for all time steps. As a result, one can set $\delta\tau'$ to be very small and obtain $\gK$ with very high precision and little numerical effort. In all our calculations, $\gK$ can be built as a GMPS within a few seconds.

\subsection{Ordering of Grassmann variables}\label{sec:ordering}

\begin{figure*}
\centering
  \includegraphics[width=\columnwidth]{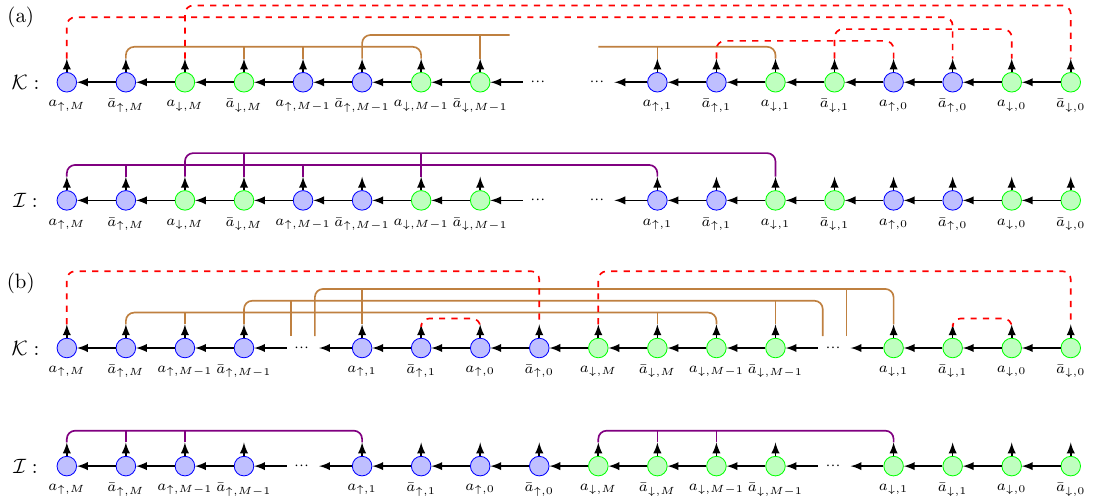} 
  \caption{The action of the interacting dynamics in $\gK$ and the influence functional $\gI$ for (a) the time-local ordering and (b) the flavor-local ordering for the single-orbital AIM. The brown solid lines represent the actions of the four-body interacting terms in the form of Eq.(\ref{eq:fourbody}), the purple solid lines represent the actions of partial IFs~\cite{ChenGuo2023}, and the red dashed lines represent the boundary actions which are responsible for the final trace in the impurity partition function. The free impurity dynamics part in $\gK$ are not shown since they are compatible with IF, namely they only act in an intra-flavor manner.
    }
    \label{fig:fig1}
\end{figure*}

The ordering of the GVs is a crucial factor that affects the performance of GTEMPO because the entanglement in an MPS can vary significantly depending on the ordering of the local basis (GVs in our case). 
The first principle of designing an appropriate ordering is that the GVs should be next to (or at least close to) their conjugates. This is because one needs to integrate out all the GVs to get a scalar output at the end of the calculation, and two GVs can only be conveniently integrated out as a pair if they are conjugate of and next to each other. 

Besides that, there are two very different strategies to order the GVs, namely the \textit{time-local} ordering and the \textit{flavor-local} ordering. For time-local ordering, the GVs at the same imaginary-time step are located in nearby positions. For flavor-local ordering, the GVs from the same flavor are located in nearby positions. The two strategies as well as the actions of the interacting dynamics and the IF are shown in Fig.~\ref{fig:fig1} for the specific case of two flavors, corresponding to the single-orbital AIM. 
The discretized interacting dynamics can be decomposed into $M$ four-body terms, exemplified as
\begin{align}\label{eq:fourbody}
e^{\lambda \abar_{\uparrow, j+1}\abar_{\downarrow, j+1}a_{\downarrow, j}a_{\uparrow, j} } = 1 + \lambda \abar_{\uparrow, j+1}\abar_{\downarrow, j+1}a_{\downarrow, j}a_{\uparrow, j}.
\end{align}
Therefore each four-body term can be written as the addition of two separable GMPSs (the unit is the vacuum state of the Grassmann space, which is itself a separable GMPS), which is thus a GMPS with bond dimension $2$.
From Fig.~\ref{fig:fig1}(a),
we can see that in the time-local ordering the four-body terms (the brown solid lines) at different times overlap at most twice. 
Therefore the bond dimension of $\gK$ will be exactly $4$ for this ordering. However, the action of IF on the two flavors completely overlaps with each other. If we assume that acting the IF on each flavor will result in a GMPS with bond dimension $\chi$, then the bond dimension of $\gI$ in Fig.~\ref{fig:fig1}(a) will be $\chi^2$.
In comparison, for the flavor-local ordering in Fig.~\ref{fig:fig1}(b), the actions of the IFs on the two flavors do not overlap, therefore the bond dimension of $\gI$ will still be $\chi$.
In the meantime, the four-body terms of the interacting dynamics completely overlap with each other, and the bond dimension of the GMPS for $\gK$ will scale as $2^M$. 
These arguments can be generalized to more flavors. In general for $n$ flavors, in the time-local ordering, if one builds the IF as a single GMPS, its bond dimension will scale as $\chi^n$, while the bond dimension of $\gK$ is a constant; in the flavor-local ordering, the IF can be built as a GMPS with bond dimension $\chi$, but the bond dimension of $\gK$ scales as $2^M$. 

Fig.~\ref{fig:fig1} thus reveals the origin of the hardness of GTEMPO: the interacting dynamics (more generally, the inter-flavor couplings) and the IF favors completely different ordering of GVs. 
One may think that time-local ordering is preferred for strong interaction and flavor-local ordering is preferred for weak interaction. However, in practice, we observe that the flavor-local ordering could only be advantageous if the interaction strength is very small compared to the rest scales (such that one can truncate the bond dimension of $\gK$ without significant loss of accuracy), and since one is usually interested in the long-time dynamics with large $M$ (the low-temperature regime), the time-local ordering is almost always preferable.
% \gcc{This situation resembles the case in CTQMC where CT-hyb~\cite{XXX} (hybridization expansion which treats $\Hhyb$ as perturbation) .
Therefore throughout this work, we will stick to the time-local ordering.

For multiple flavors, building the IF for different flavors as a single GMPS would itself be very expensive, due to the exponential scaling of the bond dimension with the number of flavors in the time-local ordering. A workaround is that one can build one MPS-IF per flavor independently, and then use the zipup algorithm proposed in Ref.~\cite{ChenGuo2023} to compute expectation values in the end. In the latter approach, one needs to build $n$ MPS-IFs in total, each with bond dimension $\chi$, which is usually very cheap. However, the computational cost of computing expectation values still grows exponentially, but with a smaller exponent as will be discussed in detail in Sec.~\ref{sec:zipup}.

Last, we mention that the red dashed lines in Fig.~\ref{fig:fig1}, which correspond to the boundary terms in Eq.(\ref{eq:K}) and are responsible for the final trace, will also increase the bond dimension of $\gK$. 
For two flavors there are $4$ such terms in total. In the time-local ordering, two of them are short-range and the rest two are long-range. This is in comparison with the PI in the real-time axis where no such boundary terms exist.
Each long-range term will increase the bond dimension of $\gK$ by a factor of $2$. For multi-orbital AIMs this increase in bond dimension due to these boundary terms becomes significant (for example $64$x for $3$ orbitals). This problem can be solved by realizing that the boundary terms are simply two-body terms similar to the form of Eq. (\ref{eq:fourbody}) (see Eq.(\ref{eq:Ktoulouse}) for example). Each two-body term, when acting on an existing GMPS with bond dimension $\chi_{\tilde{\gK}}$, results in the addition of two GMPSs, each with bond dimension $\chi_{\tilde{\gK}}$. 
For $n$ flavors, $\gK$ can be written as the addition of $2^{n}$ GMPSs resulting from splitting of the long-range boundary terms:
\begin{align}\label{eq:Kadd}
\gK = \sum_{l=1}^{2^{n}} \tilde{\gK}_l.
\end{align}
Similar to the zipup algorithm where GMPSs are multiplied on the fly, we does not need to actually perform the addition. Instead, the expectation values can be directly computed with each $\tilde{\gK}_l$ and then sum over all the results. Using this technique, for an $n_o$-orbital AIM, we can equivalently transform one calculation related to a single huge GMPS for $\gK$ with bond dimension $\chi_{\gK}$ into $2^{2n_o}$ calculations related to $2^{2n_o}$ smaller GMPSs for the $\tilde{\gK}_l$s in Eq.(\ref{eq:Kadd}), each with bond dimension $\chi_{\tilde{\gK}} = \chi_{\gK}/2^{2n_o}$. This technique will be used throughout this work, even for cases with only a single flavor or a single orbital.

\subsection{The zipup algorithm for computing expectation values}\label{sec:zipup}

\begin{figure}
\centering
  \includegraphics[]{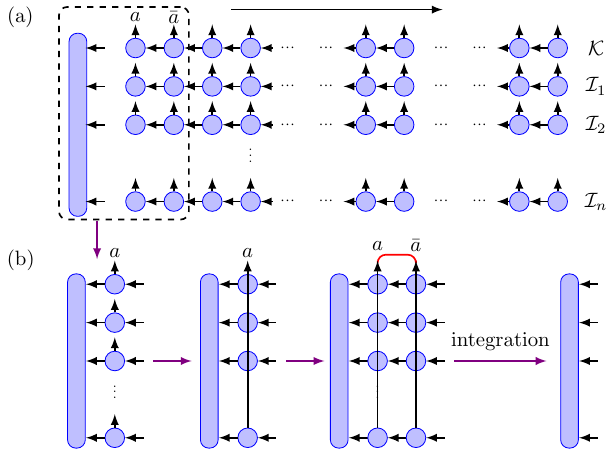} 
  \caption{(a) The zipup algorithm to integrate the Grassmann path integral with one GMPS for $\gK$ and $n$ MPS-IFs denoted from $\gI_1$ to $\gI_n$. The left bar is a trivial Grassmann tensor with a single element $1$, which is the starting point of the left-to-right sweep. (b) An elementary update during the sweep which integrates a pair of conjugate GVs denoted as $a$ and $\abar$, the final result has the same form as the input, namely a rank-$(n+1)$ Grassmann tensor.
    }
    \label{fig:fig2}
\end{figure}

In this section, we describe the zipup algorithm proposed in Ref.~\cite{ChenGuo2023} and analyze the advantage of it compared to the original TEMPO approach in detail.
In the formalism of the original TEMPO, the bare impurity dynamics and the IF are multiplied together to obtain the augmented density tensor (ADT) as a single MPS, namely
\begin{align}\label{eq:ADT}
\gA[\boldabar, \bolda] = \gK[\boldabar,\bolda]\sum_p \gI[\boldabar_p,\bolda_p].
\end{align}
Based on the ADT, one can compute multi-time correlation functions by inserting GVs at different times into the PI and integrating the resulting expression. For equilibrium QIPs, a primary class of observables of interest is the Matsubara Green's function defined as
\begin{align}\label{eq:matsubara}
G_{pq}(\tau) := \langle  \aop_{p}(\tau) \adop_{q}(0)\rangle = \langle  e^{\tau \Hop}  \aop_{p} e^{-\tau \Hop}  \adop_{q}\rangle 
= \frac{\Tr\left(e^{-(\beta-\tau) \Hop} \aop_{p} e^{-\tau \Hop} \adop_{q} \right)}{\Tr\left(e^{-\beta\Hop} \right)} .
\end{align}
In the discretized path integral formalism, $g_{pq}(j)$ at time $j\delta\tau$ can be calculated as
\begin{align}\label{eq:retarded}
G_{pq}(j)=  \Zimp^{-1}(\beta) \int\mea[\boldabar, \bolda] a_{p, j}\abar_{q, 0} \gA[\boldabar, \bolda].
\end{align}
However, directly evaluating the ADT as in Eq.(\ref{eq:ADT}) could be extremely demanding in the fermionic cases compared to the bosonic case. This is because in the fermionic case, one often needs to consider multiple orbitals and each orbital has two spins, while for the prototypical spin-boson model~\cite{LeggettZwerger1987} there is only a single bath without spin. 
In GTEMPO, the IF and $\gK$ are all built as GMPSs using a series of GMPS multiplications~\cite{ChenGuo2023}.
Assuming that the bond dimension of $\gK$ is $\chi_{\gK}$, and that there are $n$ MPS-IFs corresponding to $n$ flavors, each with bond dimension $\chi$, 
then the bond dimension of $\gA$, denoted as $\chi_{\gA}$, would scale as $\chi_{\gA}=\chi_{\gK}\chi^n$ in the worse case. As a result, each site tensor of $\gA$ could have size $2\chi_{\gK}^2\chi^{2n}$ (the factor $2$ is for the physical index). One would easily run into memory issues even for very moderate values such as $\chi=10$ and $n=4$ (two orbitals). This fact also reveals the much greater computational costs of fermionic impurity problems compared to bosonic ones.

% To numerically calculate the Grassmann PI, one can build a GMPS whose bond dimension is denoted as $\chi_{\gK}$, and build one MPS-IF for the IF acting on each band (the IF acting on each band has no overlap with each other from Eq.(\ref{eq:IF})) with bond dimension $\chi$ under an MPS truncation tolerance denoted as $\varsigma$, which means that one throws away the singular values below $\varsigma$ (the whole singular vector is normalized first) during the MPS truncation (See Ref.~\cite{Schollwock2011} for example for bond truncation of MPS using a series of singular value decompositions). 

% To alleviate this problem, we proposed a zipup algrithm to compute the expectation values without explicited building the ADT as a single huge GMPS in Ref.~\cite{ChenGuo2023}. The algorithm was demonstrated on the single-orbital Anderson models, but it can also be straightforwardly applied to any number of GMPSs. 

The central idea of the zipup algorithm is that one could arrange multiple GMPSs vertically as a quasi-two-dimensional tensor network whose physical indices at the same time step (on the same vertical line) share the same GV. Then one could perform a sweep from left to right (or equivalently from right to left) to accomplish the integration, during which the physical GVs at the same time step are multiplied together on the fly, as shown in Fig.~\ref{fig:fig2}. 
More in detail, the left-to-right sweep starts with a trivial environment tensor of rank-$(n+1)$ ($n$ for the MPS-IFs and $1$ for $\gK$) with a single element $1$ (e.g., the Grassmann vacuum). The sweep can be performed iteratively by a series of elementary updates and for each update, one integrates a pair of conjugate GVs. 
Fig.~\ref{fig:fig2}(b) shows such an elementary update, during which one needs to perform $2n$ Grassmann tensor multiplications between a large tensor with size $\chi_{\gK}\chi^{n-1}\times \chi$ and a small tensor with size $\chi\times\chi$ (the two tensor multiplications related to $\gK$ are neglected since we usually have $\chi_{\gK}\ll \chi$), therefore the computational cost of an elementary update is $O(2n\chi_{\gK}\chi^{n+1})$. Overall, one needs to perform $O(nM)$ such elementary updates during a full sweep,  and therefore the time cost of calculating a single expectation value scales as
\begin{align}\label{eq:timecost}
\timecost = O(2n^2M\chi_{\gK}\chi^{n+1}) = O(2^{n+1} n^2M\chi_{\tilde{\gK}}\chi^{n+1}), 
\end{align}
in terms of the number of floating point operations (FPOs).
We have taken the technique to split $\gK$ as in Eq.(\ref{eq:Kadd}) into account in the second equality of Eq.(\ref{eq:timecost}).
Since only one copy of the environment tensor needs to be stored, the memory cost is (neglecting the memory costs of storing the MPS-IFs and $\gK$)
\begin{align}\label{eq:spacecost}
\spacecost = O(\chi_{\tilde{\gK}}\chi^n).
\end{align}
We can see that compared to the original TEMPO which builds up a single ADT, the zipup algorithm approximately reduces the exponent in both the time and memory costs by half if $\chi_{\gA} \propto \chi_{\gK}\chi^n$. 
% Therefore the zipup algorithm will be extensively in this work used to deal with multiple-band QIPs.

Besides the multiplication of two Grassmann tensors, two additional operations are required to accomplish one elementary update. The first operation is the multiplication of several physical indices at the same time step into a single physical index, which is used in the two middle steps in Fig.~\ref{fig:fig2}(b). Mathematically, a Grassmann tensor $\gA$ in the space spanned by $n$ GVs from $\xi_1$ to $\xi_n$ can be written as~\cite{ChenGuo2023}
\begin{align}
\gA = \sum_{i_1, i_2, \dots, i_n=0}^1 A^{i_n, i_{n-1}, \dots, i_1} \xi_n^{i_n} \xi_{n-1}^{i_{n-1}} \cdots \xi_1^{i_1},
\end{align}
where the coefficient $A^{i_n, i_{n-1}, \dots, i_1}$ is a normal rank-$n$ tensor, each dimension with size $2$. If the $\xi_i$s are all the same, namely $\xi_i = \xi$, then $\gA$ can be reduced into a rank-$1$ Grassmann tensor $\gB = \sum_{i_1=0}^1 B^{i_1}\xi^{i_1}$ with
\begin{align}
  B^{0} = A^{0, 0, \dots, 0},\quad
  B^1 = \sum_{i_1+i_2+\cdots+i_n = 1} A^{i_n, i_{n-1}, \dots, i_1}.
\end{align}
The other operation is to integrate out a pair of GVs which are conjugate of each other. Assuming that $\xi_1 = \bar{\xi}_2$, then integrating out these two GVs (with the measure $\dd \bar{\xi}_1 \dd\xi_1 e^{-\bar{\xi}_1 \xi_1 }$) would result in a rank-$(n-2)$ Grassmann tensor $\gB = \sum_{i_3, \dots, i_n} B^{i_n, i_{n-1}, \dots, i_3} \xi_{n}^{i_n}\xi_{n-1}^{i_{n-1}}\cdots \xi_3^{i_3}$ with
\begin{align}
B^{i_n, i_{n-1}, \dots, i_3} = \sum_{k=0}^1 A^{i_n, i_{n-1}, \dots, i_3, k, k}.
\end{align}
These two operations together with the Grassmann tensor multiplication (for which we used the package TensorKit.jl~\cite{TensorKit}) are sufficient to implement the zipup algorithm.

\subsection{Parallelizability of GTEMPO}
To this end, we briefly discuss the parallelizability of the GTEMPO method for equilibrium QIPs at finite temperatures. Generally, if one builds one MPS-IF per flavor, then the cost of building those MPS-IFs is negligible compared to the later zipup algorithm for computing expectation values (the calculations of $n$ MPS-IFs can also be perfectly parallelized over $n$ processes). For the computationally intensive part of calculating expectation values, e.g., Eq. (\ref{eq:retarded}), there are two major sources for perfect parallelization (with essentially no data communication). The first is that one could straightforwardly parallelize the calculations of $g_{pq}(j)$ for different $p$, $q$ and $j$. The second is that once we decompose $\gK$ as in Eq.(\ref{eq:Kadd}), then the calculation becomes the addition of $2^{n}$ terms, each with exactly the same numerical cost and the calculation of each term is independent of each other. Moreover, these two sources of parallelization can be carried out simultaneously. As a result, the calculation the Matsubara Green's function for all the discrete time steps can be perfectly parallelized over $M\times 2^{n}$ processes for fixed $p$ and $q$. Taking moderate values of $M=1000$ and $n=4$, the GTEMPO method can be easily and perfectly parallelized over more than $10$ thousands of processes, which is an interesting feature since the parallelizability of tensor-network-based algorithms is often very poor compared to the Monte Carlo methods. This feature could be very helpful in the future to significantly speed up the GTEMPO calculations for equilibrium QIPs.

In the following, we will numerically demonstrate the effectiveness of the GTEMPO method on equilibrium QIPs with increasing complexity.

\section{Toulouse model}\label{sec:toulouse}

First, we study the non-interacting case, referred to as the Toulouse model~\cite{leggett1987-dynamics} or the Fano-Anderson model~\cite{mahan2000-many}, the impurity Hamiltonian of which is simply
\begin{align}
\Himp &=\varepsilon_d \adop\aop,
\end{align}
where $\varepsilon_d$ is the on-site energy. Here the flavor indices are suppressed since there is only one flavor.
The Toulouse model is ideal for us to explore the effectiveness of the GTEMPO method since 1) the results can be compared to analytic solutions; 2) the scaling of the bond dimension of the MPS-IF, in this case, is the same as the MPS-IF per flavor for higher-orbital impurity models, thus one could use the information here to accurately predict the computational costs of more complicated cases based on Eqs.(\ref{eq:timecost},\ref{eq:spacecost}).
The bare impurity dynamics of the Toulouse model can be calculated exactly:
\begin{align}\label{eq:Ktoulouse}
\gK[\boldabar, \bolda] = e^{\abar_0 a_M} e^{\eta \abar_M a_{M-1}} \cdots e^{\eta\abar_2a_1}e^{\abar_1 a_0},
\end{align}
where $\eta = e^{-\delta\tau \varepsilon_d}$. The first and last terms on the rhs correspond to the boundary terms in Eq.(\ref{eq:K}). In the time-local ordering we have $\chi_{\tilde{\gK}}=2$ ($\chi_{\gK}=4$).

The only two sources of errors in the GTEMPO method, given that $\gK$ can be calculated very precisely (either in an analytical or numerically very accurate way as shown in Sec.~\ref{sec:PI}), are the time discretization error of the PI and the bond truncation error during compression of the MPS-IF. In the following, we denote the bond truncation tolerance of the MPS-IF per flavor as $\varsigma$ (which means that we throw away the singular values whose relative weights are below $\varsigma$~\cite{Schollwock2011,ChenGuo2023}). 
For our calculations of the Matsubara Green's function $G_{pq}(\tau)$ in all cases, we set $p$ and $q$ to be the first flavor and omit them.

\begin{figure}
\centering
  \includegraphics[]{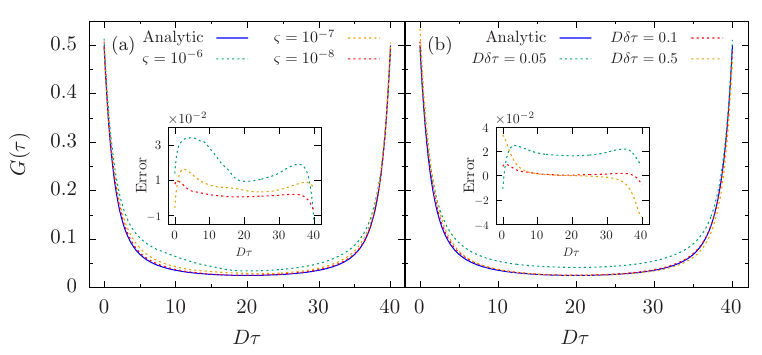} 
  \caption{Matsubara Green's function of the Toulouse model at $\varepsilon_d=0$ and $D\beta=40$ for (a) different values of $\varsigma$ and (b) different values of $\delta\tau$. The blue solid line in both panels is the analytic solution. In panel (a), the dashed lines top to bottom are GTEMPO results calculated with $\varsigma=10^{-6}, 10^{-7}, 10^{-8}$ respectively. In panel (b), the dashed lines from top to bottom are GTEMPO results calculated with $D\delta\tau=0.05, 0.5, 0.1$ respectively. The inset in both panels shows the error between the GTEMPO results and the analytic solutions. 
    }
    \label{fig:toulouse1}
\end{figure}

In Fig.~\ref{fig:toulouse1}, we study the accuracy of GTEMPO against the two hyperparameters $\delta\tau$ and $\varsigma$, where we focus on the symmetric case (half filling) with $\varepsilon_d=0$ and $\beta=40$. In Fig.~\ref{fig:toulouse1}(a), we show the GTEMPO results against the analytical solutions for different values of $\varsigma$. We can see that the accuracy improves monotonically with decreasing $\varsigma$. For $\varsigma=10^{-8}$, the differences between GTEMPO results and the analytical solutions are of the order $10^{-2}$ (less than $2\%$), and we will stick to this value of $\varsigma$ when building the MPS-IF per flavor in our rest simulations. In Fig.~\ref{fig:toulouse1}(b), we show the GTEMPO results against the analytical solutions for $\delta \tau=0.05, 0.1,0.5$. 
We can see that the case $\delta\tau=0.1$ is the most accurate of all the three cases considered. For larger $\delta\tau=0.5$, the results in the middle part (with $10\leq \tau\leq 30$) are almost as accurate as the results for $\delta\tau=0.1$, however at the boundaries, the results are clearly away from half filling. 
Interestingly, the results with a smaller $\delta\tau=0.05$ are less accurate in the middle part compared to the rest two cases. This reveals an intricate interplay between these two hyperparameters: for smaller $\delta\tau$ one generally needs to use a smaller $\varsigma$ as well. We also observe that a smaller $\delta\tau$ would result in smaller $\chi$ for the MPS-IF even under a slightly lower $\varsigma$, and generally the results at the boundaries are less accurate than those at the intermediate imaginary times. These results suggest that one may use a smaller $\delta\tau$ at the boundaries, but a much looser $\delta\tau$ in the middle to accelerate the performance while maintaining accuracy. In our following simulations, we will stick to $\delta\tau=0.1$. 

\begin{figure}
\centering
  \includegraphics[]{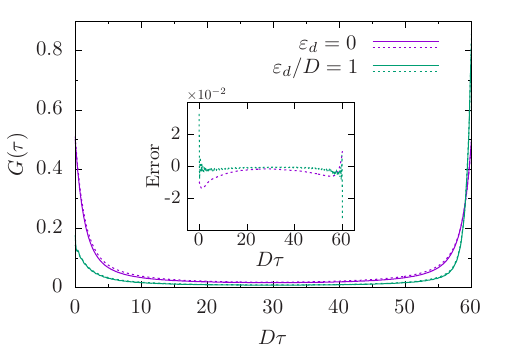} 
  \caption{Matsubara Green's function of the Toulouse model in the symmetric (with $\varepsilon_d=0$, purple lines) and non-symmetric (with $\varepsilon_d=1$, green lines) cases. The dotted lines are GTEMPO results and the solid lines are the corresponding analytical solutions. The inset shows the error between the GTEMPO results and the analytical solutions in both cases.
    }
    \label{fig:toulouse2}
\end{figure}

In Fig.~\ref{fig:toulouse2}, we study the accuracy of GTEMPO for both the symmetric and non-symmetric (with $\varepsilon_d=1$) cases with $\beta=60$. 
We can see that in the symmetric case, the errors compared to analytical solutions are still within $2\%$, despite a larger value of $\beta$. 
In the meantime, the errors in the non-symmetric case are more peaked at the boundaries compared to the symmetric case. This is because, in the non-symmetric case, the Matsubara Green's function has different values at two boundaries.

Since the IFs act on different flavors independently, for higher-orbital impurity models, if one builds one MPS-IF per flavor and computes expectation values using the zipup algorithm, then those MPS-IFs will have exactly the same bond dimension. Therefore by studying the scaling of the bond dimension $\chi$ of the MPS-IF for the Toulouse model (which has only one flavor), one can already determine the computational costs for more complicated cases (the bond dimension $\chi_{\gK}$ of $\gK$ could also increase exponentially with the number of orbitals $n_o$, however, its effect is not so significant as we only consider $n_o\leq 3$ in this work and $\chi_{\gK}$ is independent of $\beta$ in the time-local ordering). 

\begin{figure}
\centering
  \includegraphics[]{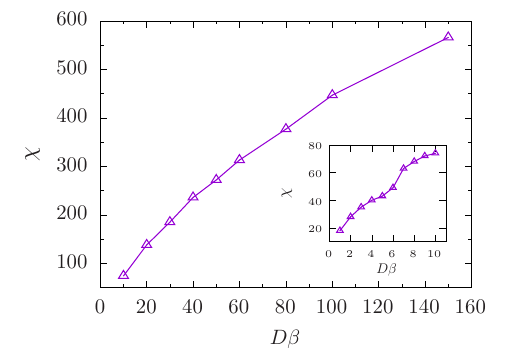} 
  \caption{Scaling of the bond dimension $\chi$ of the MPS-IF for the Toulouse model against the inverse temperature $\beta$. The inset shows the scaling of $\chi$ for small values of $\beta$.
    }
    \label{fig:toulouse3}
\end{figure}

In Fig.~\ref{fig:toulouse3} we study the scaling the $\chi$ against $\beta$, for a fixed $\varsigma=10^{-8}$. We can see that $\chi$ grows close to linearly against $\beta$. This is in sharp comparison with the calculations in the real-time axis~\cite{ChenGuo2023}, where one observes that $\chi$ increases at the beginning and quickly saturates at a value less than $20$ for the Toulouse model. The reason for the increase of bond dimension of the MPS-IF in the imaginary-time axis can be understood from the factor $e^{-\varepsilon(\tau-\tau')}$ in the hybridization function in Eq.(\ref{eq:hybridization-function}).
Due to the existence of this
factor, the correlation in the IF (e.g., the exponent in the IF in Eq.(\ref{eq:IF}))
will not simply decay with $\abs{\tau-\tau'}$ since it contains contributions of $\Delta(\tau, \tau')$ from both $\tau>\tau'$ and $\tau<\tau'$. For example, when $\varepsilon>0$, then $\Delta(\tau, \tau')$ will decay with increasing $\abs{\tau-\tau'}$ for $\tau>\tau'$, but will increase for $\tau<\tau'$. 
% The deep reason is that in the definition of the Matsubara Green's function in Eq.(\ref{eq:matsubara}), 
In comparison, for the real-time path integral, the correlation in the IF would
generally decay with the time distance $|t-t'|$, thus the bond
dimension of the MPS-IF in the real-time axis could be very moderate even for very large $t$.
%\gcc{give some brief explanations in this case together with some references?} \textcolor{blue}{maybe not necessary}.

\section{Single-orbital impurity model}\label{sec:oneorbitals}
\begin{figure}
\centering
  \includegraphics[]{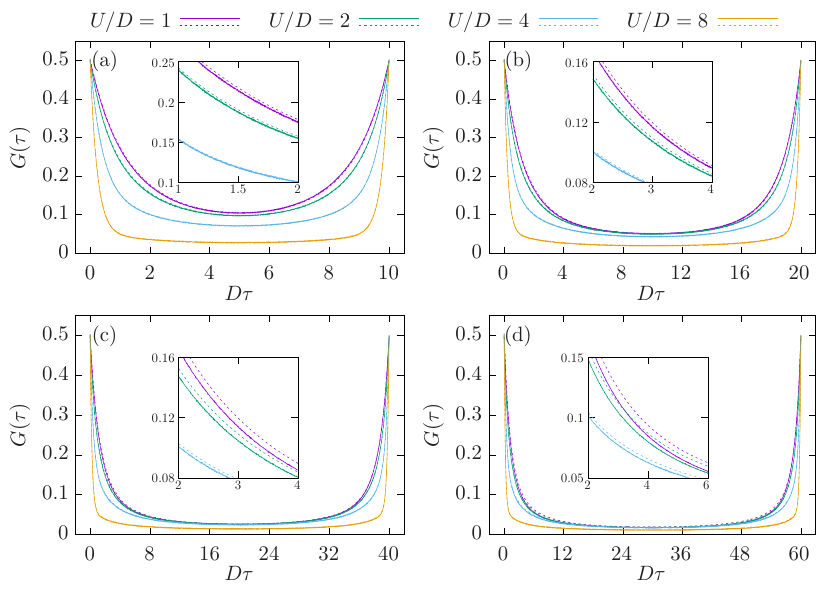} 
  \caption{Matsubara Green's function of the single-orbital Anderson impurity model for (a) $D\beta=10$, (b) $D\beta=20$, (c) $D\beta=40$ and (d) $D\beta=60$. The dashed lines in each panel from top to bottom are GTEMPO results for $U/D=1,2,4,8$ respectively, and the solid lines with the same colors are the corresponding CTQMC results. The inset in each panel shows the zooms at small values of $D\tau$.} %of the imaginary time interval $D\tau\in [1, 2]$.}
    \label{fig:oneorbital1}
\end{figure}

Now we consider the single-orbital AIM for which the impurity Hamiltonian is
\begin{align}
\Himp &= \varepsilon_d \sum_{\sigma}\adop_{\sigma}\aop_{\sigma} + U \adop_{\uparrow}\adop_{\downarrow}\aop_{\downarrow}\aop_{\uparrow},
\end{align}
where $\varepsilon_d$ is the on-site energy of the impurity and $U$ is the Coulomb interaction strength, $\sigma \in \{\uparrow, \downarrow\}$ is the spin index.
The bare impurity dynamics of this model can also be calculated exactly, with each propagator
\begin{align}\label{eq:single-orbital-exact-K}
 \propagator_{j, j-1} = e^{\eta\sum_{\sigma}\abar_{\sigma,j}a_{\sigma,j-1}}e^{\eta^2(e^{-\delta\tau U}-1)\abar_{\uparrow,j}\abar_{\downarrow,j}a_{\downarrow,j-1}a_{\uparrow,j-1}},
\end{align}
where $\eta$ is the same as the case of the Toulouse model.
For the single-orbital or higher-orbital cases, we will focus on the half-filling case, providing a quick and straightforward check for the GTEMPO results. For single-orbital AIM the half-filling condition is $\varepsilon_d = -U/2$.

In Fig.~\ref{fig:oneorbital1}, we study the accuracy of GTEMPO against CTQMC for the single-orbital AIM for $\beta=10,20,40,60$ in the four panels, respectively. For each $\beta$, we consider $U=1,2,4,8$. Here we have calculated one MPS-IF per spin species for GTEMPO. The CTQMC results are calculated using the TRIQS package~\cite{ParcolletSeth2015,SethParcollet2016}, where we have used $32$ Markov chains and generated $10^7$ samples from each chain.
We can see that the GTEMPO results agree quite well with the CTQMC results for all the cases considered. Generally, the deviations of the GTEMPO results from the CTQMC results are slightly larger for larger $\beta$, and for each value of $\beta$, the deviations become smaller for larger $U$ (The scale of $U=8$ is quite different from the rest $U$s and thus the case $U=8$ is not shown in the insets, but it can be seen that for this case the two sets of results are almost completely on top of each other). This latter feature is because 
we have chosen the time-local ordering where we can easily treat $\gK$ exactly and that the weight of $\gK$ in the ADT in Eq.(\ref{eq:ADT}) becomes larger for larger $U$ (the MPS-IF is the same for different $U$s).
%\gcc{This feature again indicates the similarity of the time-local ordering to the CT-hyb method, which works better for larger $U$} \textcolor{blue}{maybe not necessary}.

To this end, we give some explicit estimations of the computational cost of the GTEMPO method for higher-orbital AIMs. We assume that the bond dimension of the bare impurity dynamics $\gK$ scales as $\chi_{\tilde{\gK}} \propto 2^n$. From Fig.~\ref{fig:toulouse3}, we consider $\chi=74$ for $\beta=10$, $\chi = 138$ for $\beta = 20$ and $\chi=236$ for $\beta=40$. Then from Eqs.(\ref{eq:timecost},\ref{eq:spacecost}), for calculating a single expectation value of a two-orbital AIM, the time costs are $\timecost_{\beta=10} \approx 1.8\times 10^{15}$, $\timecost_{\beta=20} \approx 8\times 10^{16}$  and $\timecost_{\beta=40} \approx 2.4\times 10^{18}$ FPOs, and the memory costs are $\spacecost_{\beta=10} \approx 3.6$, $\spacecost_{\beta=20} \approx 43.5$ and $\spacecost_{\beta=40} \approx 372$ GB respectively. 
We can see that the cases $\beta=10,20$ are manageable for state-of-the-art high-performance computing systems (the newest NVIDIA H100 has higher than $100$ TFLOPS single-precision performance with $80$ GB memory), while the $\beta=40$ case is very demanding (taking into consideration that one needs to do this calculation for $O(M)$ times).

\begin{figure}
\centering
  \includegraphics[]{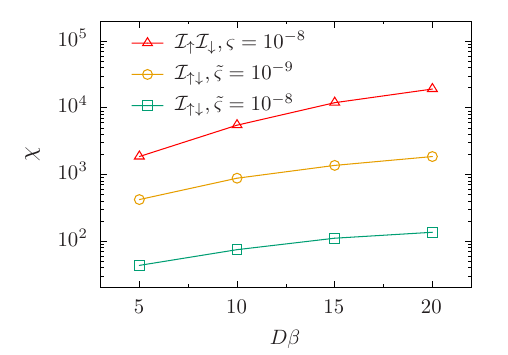} 
  \caption{Scaling of the bond dimension against the inverse temperature $\beta$, for $\gI_{\uparrow} \gI_{\downarrow}$ where the two MPS-IFs $\gI_{\uparrow}$ and $\gI_{\downarrow}$ are multiplied together without truncation (the red line with triangle), for $\gI_{\uparrow\downarrow}$ with $\tilde{\varsigma}=10^{-9}$ (the yellow line with circle) and with $\tilde{\varsigma}=10^{-8}$ (the green line with square).
    }
    \label{fig:oneorbital2}
\end{figure}

One way to further accelerate the performance of GTEMPO for multi-orbital AIMs is to combine several MPS-IFs for several flavors into a single larger GMPS, and compress the resulting GMPS using another bond truncation tolerance $\tilde{\varsigma}\leq \varsigma$ such that its bond dimension is smaller than the product of the bond dimensions of the MPS-IFs. The price to pay is that one may lose precision and the overall performance can only be evaluated for practical applications. A natural choice in this direction is to combine the two MPS-IFs for the same spatial orbital but for different spins, namely $\gI_{\uparrow}$ and $\gI_{\downarrow}$, into a single larger GMPS, denoted as $\gI_{\uparrow\downarrow}$. 
Denoting the bond dimension of $\gI_{\uparrow\downarrow}$ as $\chi_{\uparrow\downarrow}$, the time and space costs for calculating a single expectation value of an $n_o$-orbital AIM become
\begin{align}\label{eq:timecost2}
\timecost_{\uparrow\downarrow} = O(2^{n+1} n^2M\chi_{\tilde{\gK}} \chi_{\uparrow\downarrow}^{n_o+1}),
\end{align}
and 
\begin{align}\label{eq:spacecost2}
\spacecost_{\uparrow\downarrow} = O(\chi_{\tilde{\gK}} \chi_{\uparrow\downarrow}^{n_o}),
\end{align}
respectively. In Fig.~\ref{fig:oneorbital2}, we show the scaling of the bond dimensions of the GMPSs for three cases: (a) directly multiplying $\gI_{\uparrow}$ and $\gI_{\downarrow}$ together without any further bond truncation (thus the bond dimension of the resulting $\gI_{\uparrow}\gI_{\downarrow}$ is simply $\chi^2$); (b) $\gI_{\uparrow\downarrow}$ with $\tilde{\varsigma}=10^{-9}$; (c) $\gI_{\uparrow\downarrow}$ with $\tilde{\varsigma}=10^{-8}$.
We can see that compared to the base case (a), the bond dimension drops by about one order of magnitude in case (b) and by more than two orders of magnitude in case (c).

\begin{figure}
    \centering
  \includegraphics[]{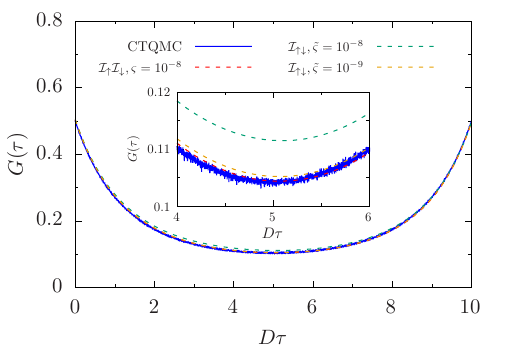} 
  \caption{Matsubara Green's function of the half-filling single-orbital AIM at $D\beta=10$ and $U/D=1$, calculated by CTQMC (the curly blue line), by GTEMPO with separate MPS-IFs (labeled by $\gI_{\uparrow}\gI_{\downarrow}$) for different spins (the red dashed line in the bottom), by GTEMPO with a joint MPS-IF for both spins (labeled by $\gI_{\uparrow\downarrow}$) under different bond truncation tolerances $\tilde{\varsigma}=10^{-8}$ (the green dashed line on the top) and $\tilde{\varsigma}=10^{-9}$ (the yellow dashed line in the middle). The inset is a zoom of the imaginary-time interval $D\tau\in [4,6]$ to better illustrate the differences among those results.
    }
    \label{fig:oneorbital3}
\end{figure}

The effectiveness of the joint MPS-IF approach is demonstrated in Fig.~\ref{fig:oneorbital3} for the half-filling single-orbital AIM at $\beta=10$ and $U=1$, where we show the Matsubara Green's function calculated by CTQMC, by GTEMPO with separate MPS-IFs for different spins (labeled by $\gI_{\uparrow}\gI_{\downarrow}$), and with a joint MPS-IF under different $\tilde{\varsigma}$ labeled by $\gI_{\uparrow\downarrow}$. We can see that the error is about $10$ percent compared to the CTQMC results with  $\tilde{\varsigma}=10^{-8}$, and becomes negligible with $\tilde{\varsigma}=10^{-9}$.
Since one expects that the accuracy requirement on the MPS-IF is less stringent for larger $U$ (see Fig.~\ref{fig:oneorbital1}), this joint MPS-IF approach is expected to be most suitable to work in the strongly interacting regime.

Now we can estimate the computational costs using Eqs.(\ref{eq:timecost2},\ref{eq:spacecost2}) under $\tilde{\varsigma}=10^{-9}$. From Fig.~\ref{fig:oneorbital2} we have $\chi_{\uparrow\downarrow} = 876, 1850$ for $\beta=10, 20$. Then for two orbitals, the time costs are $5.5 \times 10^{14}$ and $1.0\times 10^{16}$ FPOs, and the space costs are $0.09$~GB and $0.4$~GB, respectively (the storage of $\gI_{\uparrow\downarrow}$ increases significantly compared two separate $\gI_{\uparrow}$ and $\gI_{\downarrow}$, but is still moderate considering the memory of current computers and they do not need to be stored in the first-level memory, such as the GPU memory). We can see that the joint MPS-IF approach has a significant advantage in terms of the memory cost for multi-orbital AIMs. For three orbitals, the time and space costs for $\beta=10$ are already $1.7\times 10^{19}$ FPOs and $323$ GB respectively, which is likely out of reach currently.
The joint MPS-IF approach will be used for all our calculations for two and three orbitals in the following, with fixed $\tilde{\varsigma}=10^{-9}$.

\section{Multi-orbital impurity model}\label{sec:moreorbitals}

\begin{figure}
\centering
  \includegraphics[]{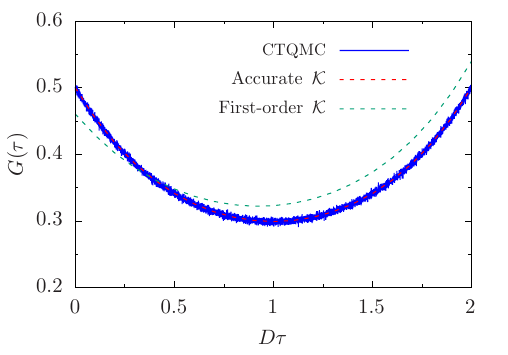} 
  \caption{Effect of the numerically accurate $\gK$ calculated using Eq.(\ref{eq:Gacc}) with $\delta\tau'=10^{-3}$ (red dashed line), compared to a first order $\gK$ with $\delta\tau=0.1$ (green dashed line). Here we have used the two-orbital AIM with impurity Hamiltonian in Eq.(\ref{eq:Himp2}). The CTQMC results (blue solid line) are also shown for reference.
    }
    \label{fig:multiorbital1}
\end{figure}

Now let us consider the multi-orbital Anderson impurity model
\cite{GullWerner2011,GeorgesMravlje2013} with the impurity Hamiltonian 
\begin{align}\label{eq:Himp2}
    \Himp=&\varepsilon_d\sum_{p}\adop_{p}\aop_{p}
    +U\sum_x\adop_{x,\uparrow}\adop_{x,\downarrow}\aop_{x,\downarrow}\aop_{x,\uparrow} \nonumber 
    +(U-2J)\sum_{x\neq y}\adop_{x,\uparrow}\adop_{y,\downarrow}\aop_{y,\downarrow}\aop_{x,\uparrow} \nonumber \\
    &+(U-3J)\sum_{x>y,\sigma}\adop_{x,\sigma}\adop_{y,\sigma}\aop_{y,\sigma}\aop_{x,\sigma}  
    -J\sum_{x\neq y}(\adop_{x,\uparrow}\adop_{x,\downarrow}\aop_{y,\uparrow}\aop_{y,\downarrow}
    +\adop_{x,\uparrow}\adop_{y,\downarrow}\aop_{y,\uparrow}\aop_{x,\downarrow}),
\end{align}
where $\varepsilon_d$ is the on-site energy, $U$ is the Coulomb interaction strength, $J$ is the Hund's coupling strength, and we have used $x, y$ for orbitals. We will consider at most three orbitals, with fixed $U=2$ and $J=0.5$. The half-filling conditions for the two-orbital and three-orbital cases are $\varepsilon_d = -(3U-5J)/2$~\cite{Sherman2020} and $\varepsilon_d = -(2.5U-5J)$~\cite{WernerMillis2009} respectively. For two orbitals the GVs within time step $j$ are arranged as $a_{1\uparrow, j}\abar_{1\uparrow, j}a_{1\downarrow, j}\abar_{1\downarrow, j} a_{2\uparrow, j}\abar_{2\uparrow, j} a_{2\downarrow, j}\abar_{2\downarrow, j}$ (the first index in the subscript is the orbital index) in the time-local ordering, and similarly for three orbitals.

For the impurity Hamiltonian in Eq.(\ref{eq:Himp2}), we are not able to derive the exact analytical expression of the propagator. Therefore we use the technique in Eq.(\ref{eq:Gacc}) to obtained numerically accurate expression for $\gK$ with $\delta\tau' = 10^{-3}$. In Fig.~\ref{fig:multiorbital1}, we show the improvement of our numerically accurate expression for $\gK$ compared to a bare first-order expression with $\delta\tau=0.1$. We can see that with the accurate $\gK$ the GTEMPO results well agree with the CTQMC results, while with the first-order $\gK$ one obtains qualitatively wrong results. 

\begin{figure}
\centering
  \includegraphics[]{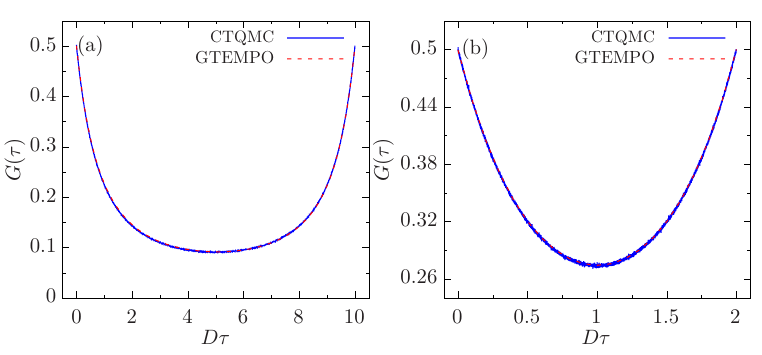} 
  \caption{Matsubara Green's function for (a) the two-orbital AIM at $D\beta=10$ and (b) the three-orbital AIM at $D\beta=2$. The red dashed line and blue solid line in both panels are the GTEMPO results and the CTQMC results respectively. 
    }
    \label{fig:multiorbital2}
\end{figure}

Finally, we study the two-orbital AIM at $\beta=10$ and the three-orbital AIM at $\beta=2$, and the results are shown in Fig.~\ref{fig:multiorbital2}(a, b) respectively. We can see that the GTEMPO results in both cases agree very well with the CTQMC results, which demonstrates the flexibility of the GTEMPO method. The calculations in Fig.~\ref{fig:multiorbital2} are already very demanding. For example, in Fig.~\ref{fig:multiorbital2}(a), we have used $10$ threads (each with $2.9$ GHz frequency) for around one week, where we have parallelized the calculations of the Matsubara Green's function at different time steps. Nevertheless, utilizing nowadays high performance computing techniques, we vision that for two-orbital AIMs one could reach $\beta=20$, but for the three-orbital model, it would still be very demanding for larger $\beta$ (see our analysis at the end of Sec.~\ref{sec:oneorbitals}).

\section{Summary}\label{sec:summary}
In summary, we have systematically studied the effectiveness of the recently proposed GTEMPO method by computing the Matsubara Green's function of equilibrium quantum impurity problems at finite temperatures. Several techniques are proposed to accelerate the performance of GTEMPO for finite-temperature calculations, including the technique to deal with the boundary condition in the impurity partition function, ordering of Grassmann variables, as well as the joint MPS-IF approach. Our numerical calculations reach $\beta=60$ for the single-orbital AIM,  $\beta=10$ for the two-orbital AIM and $\beta=2$ the three-orbital AIM, and in all cases, we show good agreement of the GTEMPO results with the CTQMC calculations. Based on rigorous estimations of the computational costs, we vision that $\beta=20$ could be reached for the two-orbital AIM with nowadays high-performance computing techniques, while for high-orbital AIMs only very high-temperature regimes could be explored under the current formalism of GTEMPO. Our work thus paves the way to apply GTEMPO as an imaginary-time impurity solver.
%
% =======================  Acknowledgements  ======================
%

\section*{Acknowledgements}
This work is supported by the National Natural Science Foundation of China under Grant No. 12104328 and 12305049. C. G. is supported by the Open Research Fund from State Key Laboratory of High Performance Computing of China (Grant No. 202201-00).
\vskip10pt

\bibliographystyle{iopart-num}
\bibliography{refs}

\end{document}